\documentclass[sigconf,nonacm]{acmart}

\AtBeginDocument{%
  }

\copyrightyear{2025}
\acmYear{2025}
\setcopyright{cc}
\setcctype{by}
\acmConference[FAccT '25]{The 2025 ACM Conference on Fairness, Accountability, and Transparency}{June 23--26, 2025}{Athens, Greece}
\acmBooktitle{The 2025 ACM Conference on Fairness, Accountability, and Transparency (FAccT '25), June 23--26, 2025, Athens, Greece}\acmDOI{10.1145/3715275.3732115}
\acmISBN{979-8-4007-1482-5/2025/06}

\usepackage{enumitem}
\usepackage{multirow}
\usepackage{graphicx}
\usepackage{caption}
\usepackage{subcaption}

\begin{document}

\title{Ties of Trust: a bowtie model to uncover trustor-trustee relationships in LLMs}

\author{Eva Paraschou}
\email{eparascho@csd.auth.gr}
\affiliation{%
  \institution{Aristotle University of Thessaloniki, School of Informatics}
  \city{Thessaloniki, 54124}
  \country{Greece}
}

\author{Maria Michali}
\email{mmichals@csd.auth.gr}
\affiliation{%
  \institution{Aristotle University of Thessaloniki, School of Informatics}
  \city{Thessaloniki, 54124}
  \country{Greece}
}

\author{Sofia Yfantidou}
\email{syfantid@csd.auth.gr }
\affiliation{%
  \institution{Aristotle University of Thessaloniki, School of Informatics}
  \city{Thessaloniki, 54124}
  \country{Greece}
}

\author{Stelios Karamanidis}
\email{skaraman@csd.auth.gr }
\affiliation{%
  \institution{Aristotle University of Thessaloniki, School of Informatics}
  \city{Thessaloniki, 54124}
  \country{Greece}
}

\author{Stefanos Rafail Kalogeros}
\email{stefkalo@csd.auth.gr }
\affiliation{%
  \institution{Aristotle University of Thessaloniki, School of Informatics}
  \city{Thessaloniki, 54124}
  \country{Greece}
}

\author{Athena Vakali}
\email{avakali@csd.auth.gr}
\affiliation{%
  \institution{Aristotle University of Thessaloniki, School of Informatics}
  \city{Thessaloniki, 54124}
  \country{Greece}
}

\renewcommand{\shortauthors}{Paraschou et al.}

\begin{abstract}
    The rapid and unprecedented dominance of Artificial Intelligence (AI), particularly through Large Language Models (LLMs), has raised critical trust challenges in high-stakes domains like politics. Biased LLMs' decisions and misinformation undermine democratic processes, and existing trust models fail to address the intricacies of trust in LLMs. Currently, oversimplified, one-directional approaches have largely overlooked the many relationships between trustor (user) contextual factors (e.g. ideology, perceptions) and trustee (LLMs) systemic elements (e.g. scientists, tool's features). In this work, we introduce a bowtie model for holistically conceptualizing and formulating trust in LLMs, with a core component comprehensively exploring trust by tying its two sides, namely the trustor and the trustee, as well as their intricate relationships. We uncover these relationships within the proposed bowtie model and beyond to its sociotechnical ecosystem, through a mixed-methods explanatory study, that exploits a political discourse analysis tool (integrating ChatGPT), by exploring and responding to the next critical questions: 1) How do trustor's contextual factors influence trust-related actions? 2) How do these factors influence and interact with trustee systemic elements? 3) How does trust itself vary across trustee systemic elements? Our bowtie-based explanatory analysis reveals that past experiences and familiarity significantly shape trustor's trust-related actions; not all trustor contextual factors equally influence trustee systemic elements; and trustee's human-in-the-loop features enhance trust, while lack of transparency decreases it. Finally, this solid evidence is exploited to deliver recommendations, insights and pathways towards building robust trusting ecosystems in LLM-based solutions.
\end{abstract}




\maketitle

\section{Introduction}\label{introduction}
Artificial Intelligence (AI), with its Generative Large Language Models (LLMs) flagship, has rapidly taken the world by storm. Trust in AI and LLMs remains fragile and brittle: 56\% of people express strong concerns about AI bias, privacy and data protection \cite{ipsos2023}; public skepticism about AI has significantly raised \cite{aiindex2024}, largely due to LLMs bias \cite{StanfordHAI2024} and ``hallucinated'' content which fuels misinformation \cite{Church_2024}. Relying on LLMs for decision-making demands for uncharted escapes from misinformation, online manipulation, propaganda, and polarization \cite{barman2024dark, shah2024navigating, ferrara2017disinformation}. Politics are critically impacted, since LLMs are heavily used for content generation, opinion shaping, and public discourse. Political discourse and democratic values are threatened by LLMs and AI tools used by political-authoritative stakeholders, due to their opaqueness, unfairness and human oversight faults \cite{verma2019weapons}.
%
\textbf{Redefining trust in LLMs} unprecedented and evolving reality is critical, even when viewed through the AI-tailored definition of trust: ``\textit{attitude that an agent will help achieve an individual’s goals in a situation characterized by uncertainty and vulnerability}'' \cite{lee2004trust}, as it oversimplifies the complexity of trust in adaptive, dynamically changing environments \cite{hoff2015trust, schaefer2016meta}. To redefine trust in LLMs, we must carefully consider the many complex and newly-emerged inter-relationships between the human side i.e. \textit{trustor} (in our case user placing their trust), and the agent-side i.e. the \textit{trustee} (in our case LLMs object/technology being trusted). The multiplicity and deviation of trustor-side contextual factors (e.g., demographics, perceptions) and the trustee-side systemic elements (e.g. scientific discipline, scientists), demand for a new evidence-based and pragmatic model to overcome research and implementation trust in LLMs \textit{\textbf{G}}aps.

Firstly, \textbf{trust is conceptualized by oversimplified models (\textit{G1})} across disciplines, such as Mayer et al.'s \cite{mayer1995integrative} in organization studies, Ghosh's \cite{ghosh2001student} in institutions, Barber \cite{barber1983logic} in sociology, and Holton's \cite{holton1994deciding} in philosophy, following a reductionist view not suitable for LLMs. AI trustworthiness was studied using Mayer et al's \cite{kim2024m, kim2023humans} and AI-based academic prediction with Ghosh’s models \cite{lunich2024explainable}, both focusing on trustee-side elements (benevolence, ability; transparency, reliability), neglecting trustor-side contextual factors. However, when trust in LLMs is challenged in politics, personal beliefs and systemic vulnerabilities must not be overlooked.
%
Secondly, \textbf{limited exploration of the intra- and inter-relationships among trustor and trustee sides (\textit{G2})}, is evident in earlier models \cite{mayer1995integrative, ghosh2001student} which adopt a one-directional ``trustee to trustor'' approach, neglecting the many trustor contextual factors and their correlations and adjustments with more trustee systemic elements. Such models limit our understanding of how and why trust in LLMs erodes, especially to political domain trustor-side (e.g. voters) who remains largely skeptical and distrust LLMs (e.g. political discourse tools) opaqueness \cite{tenove2018digital}. 
%
Finally, \textbf{the influence that LLMs human-centric elements have on trust is still under-explored (\textit{G3})}, since scholars partially examine how trustor's involvement, who are mostly assigned a role at LLMs' evaluation, affects their trust in LLMs \cite{honeycutt2020soliciting, wang2022human, sutton2018digitized}. To the best of our knowledge, studies which question and report the suitability of human-in-the-loop (HITL) approaches integrated in the trustee (in our case LLMs) and the ways they influence trustor's particular attitudes, are largely missing. However, in politics, stakeholders' authoritative role (trustee-side) embedded in HITL approaches may significantly affect trust in LLMs-based tools.

To resolve such critical gaps, we approach trust in LLMs as a complex sociotechnical topic, within the broader trust ecosystem where LLMs operate. In summary, we identify trustor-side contextual factors and the trustee-side systemic elements, and we introduce a particular model which redefines and enables LLMs trust exploration at both trustor-trustee sides, individually and in conjunction, based on a core deep explanatory study which reveals user (trustor) trust attitudes towards LLMs (trustee) and relationships among their contextual factors and systemic elements. Our novel approach disrupts the ``blind'' one-directional trustor-trustee exploration and contributes as next: 
\textbf{(C1) Introduce the bowtie model of trust in LLMs} by identifying the specific factors and elements of the trustor and trustee sides, inspired by contemporary frameworks for conceptualizing trust \cite{university_of_tartu_2024, jacovi2021formalizing}. Our proposed robust bowtie-shaped model goes beyond the oversimplified notions of trust in LLMs (\textit{\textbf{G1}}), having a core component to comprehensively explore trust, and two sided components to unfold users (trustors) and LLMs (trustee) many factors and elements.  
\textbf{(C2) Uncover the particular relationships within the complex sociotechnical topic of trust in LLMs}, via our bow-tie core which embeds a mixed-methods explanatory study and reveals specific intra- and inter-relationships between and within users contextual factors (trustor-side) and LLMs systemic elements (trustee-side), overcoming the limitations of isolated ad-hoc approaches (\textit{\textbf{G2}}). By exploiting an LLM- and HITL-based demo tool, we systematically collect rich quantitative and qualitative data, further employing correlation and deductive thematic analysis; 
\textbf{(C3) Deliver systematic and evidence-based knowledge and insights to researchers, policymakers, practitioners, and the general public}, through our bowtie that is proposed as an exemplar model which largely facilitates the formulation and exploration of the new and under-explored human-centric evidence of trust in LLMs (\textit{\textbf{G3}}). Our bowtie's extensible structure allows for integrating fine-grained concepts in complex, evolving, and dynamic sociotechnical contexts, and establishes a new norm for discovering distinctive and evidence-based \textbf{Ties of Trust in LLMs}.

\section{Related Work}\label{relatedwork}
Trust in AI and LLMs has been studied by adapting trust models from various domains. For example, Mayer et al.'s model \cite{mayer1995integrative}, adapted to AI, examines trustee-side trustworthiness beliefs (ability, benevolence, integrity) and trustor-side intentions (willingness to rely on the system), as applied to explore trust in computer vision applications \cite{kim2023humans} and recommender systems \cite{radensky2023think}. K\"orber’s model \cite{korber2019theoretical}, integrating Mayer et al's. \cite{mayer1995integrative} and Lee and See's \cite{lee2004trust} models, was used to study trust repair mechanisms \cite{pareek2024trust} and students’ trust in generative AI tools \cite{pareek2024trust, amoozadeh2024trust}. Ghosh’s model \cite{ghosh2001student} was used to measure how trustee-side attributes (sincerity, expertise, congeniality) affect user trust in decision trees for academic performance prediction \cite{lunich2024explainable}. McKnight et al.’s model \cite{HARRISONMCKNIGHT2002297}, combining trustee's trustworthiness beliefs and institution-based trust, and trustor's intentions and dispositions, was used to explore how uncertainty expression affects trust in LLMs \cite{kim2024m}. The HAII-TIME model \cite{pennycook2021psychology}, adapted for AI, includes trustee cues (i.e., design) that trigger cognitive heuristics and actions (i.e., engagements) that reshape trust, and was employed to explore prompt coaching and trust \cite{chen2024your}. Jacovi et al.’s model \cite{jacovi2021formalizing}, emphasizing trustor vulnerability and anticipation of AI decision impacts, was used to explore AI-trust-explainability relationships in clinical applications \cite{ferrario2022explainability}. The MATCH model \cite{liao2022designing}, integrating Mayer et al.'s \cite{mayer1995integrative}, Jacovi et al.'s \cite{jacovi2021formalizing}, and Lee and See's \cite{lee2004trust} theories, examined how online communities shape developers' trust in AI tools. Finally, Chen et al. \cite{chen2023ai} combine cognitive, affective, and behavioral trust aspects to study how labeling quality affects user trust in AI. It is evident and well-justified in \cite{gulati2024trust, benk2024twenty} that most (if not all) of the above trust models \textit{fail to capture the dynamic and multidimensional nature of trust}, as they tend to focus exclusively on either the trustor or the trustee, or examine one-directional relationships between them \textbf{\textit{(G1, G2)}}.

\begin{figure*}[htp]
    \centering
    \includegraphics[width=0.7\textwidth]{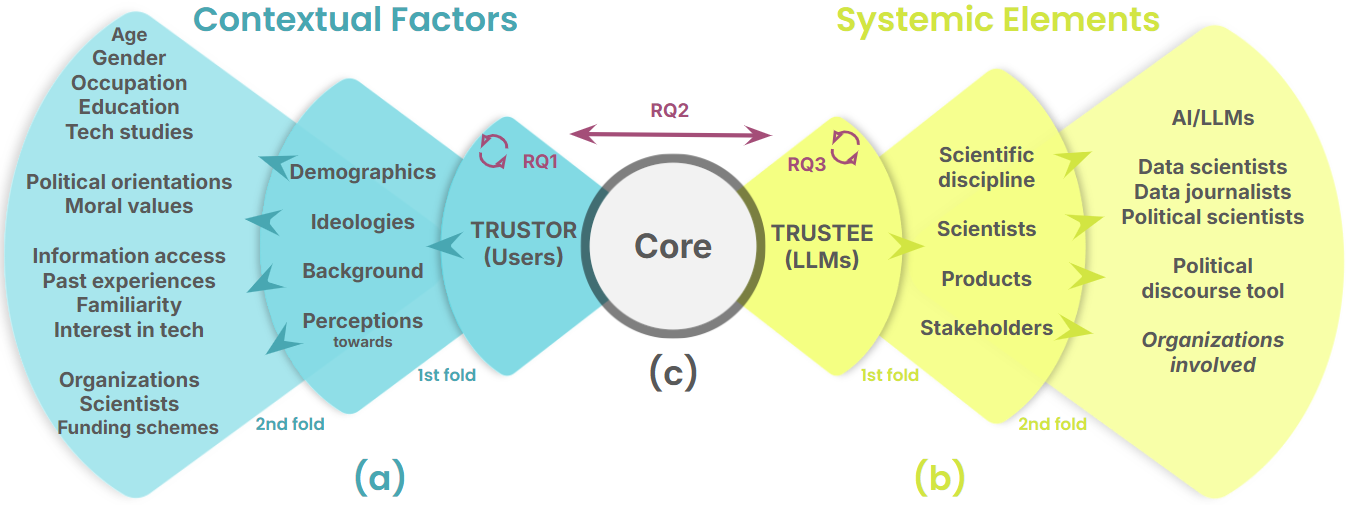}
    \caption{The bowtie model of trust in LLMs.}
    \Description{The bowtie model of trust in LLMs is structured into three main components: (a) a trustor component, encapsulating user contextual factors; (b) a trustee component, assembling LLMs systemic elements; and (c) a core component, tying the two sides. On the left, user factors are categorized into Demographics; Ideologies; Background; and Perceptions; which further unfold into the following contextual factors: age, gender, occupation, education, tech studies, political orientations, moral values, information access, past experiences, familiarity, interest in tech, institutions, scientists, funding schemas. On the right, LLMs generic systemic elements are: Scientific Discipline; Scientists; Products; and Stakeholders; which are further adapted into: AI/LLMs; data scientists, political scientists and data journalists; political discourse tool; and organization involved. The Core component links these sides and facilitates exploration of relationships.}
    \label{fig:bowtie}
\end{figure*}

Furthermore, several key trustor-side contextual factors and trustee-side systemic elements have been identified in prior studies using the above trust models or well-established questionnaire-based methods. For instance, on the trustor-side, multiple demographics affect trust levels, such as ``age'' which shapes differing preferences on anthropomorphism (attribution of human characteristics to non-human entities) \cite{inie2024ai}, ``country'' and ``culture'' in facial recognition systems \cite{ullstein2024attitudes}, ``gender'' and ``education level'' in generative AI tools  \cite{amoozadeh2024trust}. Additionally, ``domain knowledge'', ``past experiences'', ``familiarity'' and ``AI literacy'' significantly influence trust in several applications (e.g., computer vision \cite{kim2023humans}, automated decision-making \cite{schoeffer2022there}, and ChatGPT for learning \cite{bikanga2024helps}). Trustor's ``moral values'' also influence trust positively when aligning with developers’ ethical and participatory practices in AI assistants \cite{manzini2024should} and certification labels \cite{scharowski2023certification}, and especially when issued by independent organizations. Finally, positive ``trust intentions'' in automation are correlated with higher trust in AI \cite{pareek2024trust}. On the trustee-side, ``transparency'' is one of the most impactful features for increasing trust, evident across multiple application domains (e.g., AI assistants \cite{manzini2024should}, code generation tools \cite{wang2024investigating}, recommender systems \cite{radensky2023think}, generative AI tools \cite{amoozadeh2024trust}, automated decision-making \cite{schoeffer2022there}, and AI agents \cite{mehrotra2024integrity}). ``Competence'' and ``alignment'' with user goals  significantly increase trust in AI assistants \cite{manzini2024should}, while ``ease of use'' improves trust in ChatGPT \cite{ding2023students}. Broader ``attitudes about AI'' (i.e. epistemic validity and capabilities perceptions) can influence trust in facial recognition systems \cite{ullstein2024attitudes}. ``Certification'' labels significantly improve trust \cite{scharowski2023certification}, especially if issued by ``trusted organizations`` \cite{vereschak2024trust}, while institution and developer ``reputation'' also increase trust \cite{kim2023humans, lunich2024explainable}. Application-specific features like LLMs’ ``uncertainty'' \cite{kim2024m} and ``regret'' expressions \cite{pareek2024trust} influence trust levels, though anthropomorphism decreases trust in ChatGPT \cite{inie2024ai, ding2023students}. Clearly, the above findings remain mostly one-directional, from a trustor's contextual factor to a trustee's systemic element or vice versa \textbf{\textit{(G2)}}, while human-centric elements are explored only in terms of transparency \textbf{\textit{(G3)}}. Inspired by recent and holistic models, we next propose a robust bowtie-shaped model of trust in LLMs that addresses both trustor and trustee with their multiple contextual factors and systemic elements to explore their multi-directional relationships.

\section{The Bowtie of Trust in LLMs}\label{conceptualframework}
To remodel earlier trust in AI and LLMs models' monolithic approaches (\textit{\textbf{G1, G2}}), we adopt a holistic view for conceptualizing trust in LLMs. Inspired by recent and contemporary frameworks \cite{university_of_tartu_2024, jacovi2021formalizing}, which approach trust as a two-place (trustee-trustor) ``form'', we position user (trustor) contextual factors (e.g., past experiences, ideologies) on one side, and LLMs (trustee) systemic elements (e.g., scientific discipline, involved scientists) on the other. However, as trust in LLMs should systematically explore the intra- and inter-relationships among the two sides, we propose a pragmatic and robust bowtie-based model to assemble and amalgamate our two-sides (trustor-trustee) relationships (\textbf{C1}). Bowtie structures have been recently applied to automation bias detection (as the core), with causes and consequences (on the two sides) to relate and inform AI healthcare systems \cite{abdelwanis2024exploring}, and to human control (as the core) in autonomous systems, with control attributes and system's accountability (on the two sides) \cite{flemisch2023towards} to explore their relationships. To the best of our knowledge, this is the first time that bowtie is proposed to redefine, remodel, systematically frame, and study the complex sociotechnical topic of trust in LLMs.

\begin{figure*}[htp]
    \centering
    \includegraphics[width=0.9\textwidth]{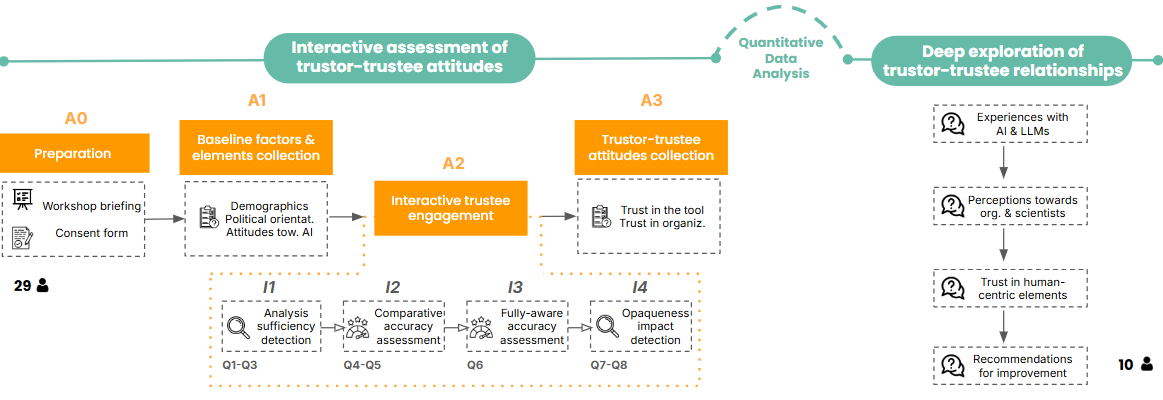}
    \caption {The bowtie core methodology and study outline.}
    \Description{The study's outline follows a sequential with (a) on-site interactive activities and (b) virtual semi-structured interviews. The interactive session contains four activities: (A0) Preparation, with workshop briefing and consent form; (A1) Baseline factors and elements collection, with demographics, political orientation, and attitudes toward AI questionnaires; (A2) Interactive trustee engagement, with four particular interactions: analysis sufficiency detection, comparative accuracy assessment, fully-aware accuracy assessment, and opaqueness impact detection; and (A3) Trustor-trustee attitudes collection, with trust in the tool and trust in organizations and scientists questionnaires. The virtual session is based on individual interviews, each consisting of four key parts: experiences with AI and LLMs; perceptions towards organizations and scientists; trust in human-centric elements; and recommendations for improvement.}
    \label{fig:study_flow}
\end{figure*}

Our bowtie model of trust in LLMs (Fig. \ref{fig:bowtie}) enables a holistic view of trust by encapsulating: (a) a trustor component with user contextual factors; (b) a trustee component, assembling LLMs systemic elements; (c) a core component (detailed in \ref{methodology}) to tie the two sides and allow for a fine-grained understanding of intra- and inter-relationships among them. Both sides unfold from broader notions to a systematic and intricate exploration of specific contextual factors and systemic elements. At the trustor component, prominent philosophical- \cite{university_of_tartu_2024} and evidence-based (see \ref{relatedwork}) factors define a generic first fold with contextual categories (Demographics; Ideologies; Background; Perceptions), which are detailed into specific contextual factors (i.e. education level, familiarity, past experiences) in the second fold. At the trustee component, the first fold includes the generic state of the art \cite{university_of_tartu_2024} elements (Scientific discipline; Scientists; Products; Stakeholders), which then unfold to the trust in LLMs context-specific systemic elements (AI/LLMs; data scientists, political scientists, data journalists; a political discourse LLM-based tool; the organizations hosting the scientists and solutions in the present study). Finally, our bowtie core acts as the trustor-trustee connecting force as it reveals knowledge and captures multi-directional relationships among the two sides: (1) the intra-relationship within trustor contextual factors; (2) the bidirectional inter-relationship between trustor contextual factors and trustee systemic elements; (3) the intra-relationship within trustee systemic elements. Our ties of trust bowtie structure offers a systematic and extensible model for exploring trust in LLMs, in response to the following critical research questions (RQs): [\textbf{RQ1}] How do trustor contextual factors influence their trust-related actions?; [\textbf{RQ2}] What is the mapping between trustor contextual factors and trustee: [\textbf{\textit{RQ2A}}] systemic elements? or [\textbf{\textit{RQ2B}}] human-centric systemic elements?; [\textbf{RQ3}] How do trustor trust attitudes differentiate over the trustee systemic elements? By addressing these RQs, we aim to untie trustor-trustee bottlenecks and offer key insights into the complex sociotechnical topic of trust in LLMs, particularly in the high-stake politics domain. Leveraging our bowtie model, we explore all possible intra- and inter-relationships within it (\textbf{C2}) and deliver knowledge and insights to researchers, policymakers, practitioners, and the general public (\textbf{C3}).

\section{The Bowtie Core: An explanatory study}\label{methodology}
At the core of our bowtie model, we position our robust methodology which enables a mixed-methods explanatory study (outlined in Fig. \ref{fig:study_flow}) to: (a) assess trustor-trustee attitudes and relationships (by on-site interactive activities), employing quantitative statistical analysis; (b) further uncover causal reasons underlying trustor-trustee attitudes and relationships (via online semi-structured interviews), using qualitative deductive thematic analysis. The entire core study has been approved by the authors' Institutional Review Board with protocol no. 9995/2024, detailed in the next subsections.

\subsection{Interactive assessment of trustor-trustee attitudes and relationships}\label{workshop}
Under the bowtie core, we adopted a ``within-subjects'' design, complemented by ``between-subjects'' tasks during the interactive on-site activities. The aim has been to identify changes in trustor-trustee attitudes under varying conditions and compare detected intra- and inter-relationships between distinct groups \cite{montoya2023selecting}. 

\subsubsection{Participants and recruitment} We recruited 29 participants using purposive sampling to ensure variation across key demographic contextual factors \cite{cooper2014business}. The sample size aligns with similar notable studies in the community employing on-site interactions and individual activities \cite{wang2024investigating, kim2023humans, radensky2023think, cheng2024would, ding2023students, vereschak2024trust}.
Participants were purposefully selected based on gender (Man: 58.62\%, Woman: 41.37\%), age (18-24: 62.06\%, 25-39: 31.03\%, 40-54: 6.89\%) occupational (Self-employed: 6.89\%, Salaried Employee: 27.58\%, Unemployed: 20.68\%, Student: 44.82\%) and educational (University students: 75.86\%, BSc / MSc / PhD holders: 24.15\%) status, through snowballing of personal and academic networks. Although familiarity with AI and LLMs was not an inclusion criterion, all participants had a (high-school or university) background in science/technology. All participants received a certificate to be rewarded for their participation. 

\begin{table*}[ht!]
\centering
\scriptsize
\caption{Questionnaires used and data collected during the on-site interactive activities, and their alignment with the bowtie model of trust in LLMs.}
\label{tab:workshop_questions}
\begin{tabular}{lll|c|c|l|l}
\toprule
\multicolumn{2}{c|}{\textbf{}} &
  \textbf{Question(naire)} &
  \textbf{Bowtie side} &
  \textbf{\begin{tabular}[c]{@{}c@{}}1st fold category\end{tabular}} &
  \textbf{\begin{tabular}[c]{@{}c@{}}2nd fold factor or element\end{tabular}} &
  \textbf{Values} \\ \toprule \toprule
\multicolumn{2}{l|}{\multirow{9}{*}{\textbf{A1}}} &
  \multirow{7}{*}{Demographics questionnaire \cite{eurobarometer_2022}} &
  \multicolumn{1}{l|}{\multirow{7}{*}{Contextual factor}} &
  \multirow{5}{*}{Demographics} &
  Gender & Woman, Man, Other \\ \cline{6-7} 
\multicolumn{2}{l|}{} & &
  \multicolumn{1}{l|}{} &  &
  Age & 18-24, 25-39, 40-54, Other \\ \cline{6-7} 
\multicolumn{2}{l|}{} &  & \multicolumn{1}{l|}{} &  &
  Occupation &
  Categorical values \\ \cline{6-7} 
\multicolumn{2}{l|}{} & & \multicolumn{1}{l|}{} & &
  Education & 15-, 16-19, 20+, still studying \\ \cline{6-7} 
\multicolumn{2}{l|}{} & & \multicolumn{1}{l|}{} & &
  Studies in science \& tech &
  University, school, both \\ \cline{5-7} 
\multicolumn{2}{l|}{} & &  \multicolumn{1}{l|}{} &
  \multirow{2}{*}{Background} &
  Information access & Categorical values \\ \cline{6-7} 
\multicolumn{2}{l|}{} & & & &
  Interest in science \& tech &
  5-Likert scale \\ \cline{3-7} 
\multicolumn{2}{l|}{} &
  Political attitudes questionnaire \cite{smith2013civic} &
  Contextual factor &
  Ideologies &
  Political orientations &
  Yes/No; 3-Likert scale; party \\ \cline{3-7} 
\multicolumn{2}{l|}{} &
  Attitudes towards AI questionnaire \cite{schepman2020initial, grassini2023development} &
  Systemic element &
  Scientific discipline &
  Attitudes towards AI &
  5-Likert scale \\ \midrule \midrule
\multicolumn{1}{l|}{\multirow{10}{*}{\textbf{A2}}} & 
\multicolumn{1}{l|}{\multirow{3}{*}{\textit{I1}}} & 
Q1: Did you press the button? \textbackslash{}\cite{kim2024m} 
& \multirow{10}{*}{Systemic element} & \multirow{10}{*}{Products} & \multirow{10}{*}{Political discourse tool} & Yes/No 
\\ \cline{3-3} \cline{7-7} 
\multicolumn{1}{l|}{} & \multicolumn{1}{l|}{} & Q2: Why yes/no?  &  &  &  & Open-ended  \\ \cline{3-3} \cline{7-7} 
\multicolumn{1}{l|}{} & \multicolumn{1}{l|}{} & Q3: Did you read the speech? \textbackslash{}\cite{kim2024m} &  &  &  & Yes/No 
\\ \cline{2-3} \cline{7-7} 
\multicolumn{1}{l|}{} & \multicolumn{1}{l|}{\multirow{2}{*}{I2}} & \multirow{2}{*}{\begin{tabular}[c]{@{}l@{}}Q4-Q5: I believe the results \\ given in this chart are accurate\end{tabular}} &  &   &  & \multirow{7}{*}{5-Likert scale} \\
\multicolumn{1}{l|}{} & \multicolumn{1}{l|}{}  & & & & & \\ \cline{2-3}
\multicolumn{1}{l|}{} & \multicolumn{1}{l|}{\multirow{2}{*}{\textit{I3}}} & \multirow{2}{*}{\begin{tabular}[c]{@{}l@{}}Q6: I believe the results produced by ChatGPT \\ and validated by data journalists are accurate\end{tabular}} & & & & \\
\multicolumn{1}{l|}{} & \multicolumn{1}{l|}{} &  &  &  &  &  \\ \cline{2-3} \multicolumn{1}{l|}{}  & \multicolumn{1}{l|}{\multirow{2}{*}{\textit{I4}}} & \multirow{2}{*}{\begin{tabular}[c]{@{}l@{}}Q7-Q8: I believe the results produced by ChatGPT \\ and validated by data journalists are accurate\end{tabular}} &  &  &  &  \\  \multicolumn{1}{l|}{}  & \multicolumn{1}{l|}{} & & & & & \\ \cline{2-3}
\multicolumn{1}{l|}{} & \multicolumn{2}{l|}{Q9: I would use such a political discourse exploration tool again \textbackslash{} \cite{tsai2021exploring}}  &  &  &   &  \\ \midrule \midrule
\multicolumn{2}{l|}{\multirow{5}{*}{\textbf{A3}}} &
  Trust in the tool questionnaire \cite{radensky2023think} &
  Systemic element &
  Products &
  Political discourse tool &
  7-Likert scale \\ \cline{3-7} 
\multicolumn{2}{l|}{} &
  \multirow{4}{*}{Trust in organizations questionnaire \cite{funk2019trust}} &
  \multirow{4}{*}{Contextual factor} &
  Ideologies &
  Moral values &
  6-Likert scale \\ \cline{5-7} 
\multicolumn{2}{l|}{} & & &
  \multirow{3}{*}{Perceptions} &
  towards organizations &
  3-Likert scale \\ \cline{6-7} 
\multicolumn{2}{l|}{} & & & &
  towards scientists &
  3-Likert \& 5-Likert scales \\ \cline{6-7} 
\multicolumn{2}{l|}{} & & & &
  towards funding scheme &
  6-Likert scale \\ \bottomrule
\end{tabular}
\end{table*}

\subsubsection{Activities procedure} 29 participants completed the activities in person in the authors' affiliated premises. These activities collected data on trustor (users/participants) contextual factors and trustee (LLMs) systemic elements, as proposed in our bowtie model of trust in LLMs (Fig. \ref{fig:bowtie}), with all question(naire)s and tasks tailored to this goal (Tab. \ref{tab:workshop_questions}). The \textbf{A}ctivities are: 

\textbf{(A0) Preparation}. Participants were briefed on the activities' purpose and completed a consent form to confirm their informed participation. They were introduced to a political discourse tool developed earlier by the authors' team to analyze pre-election political speeches in Greece\footnote{\url{https://lab.imedd.org/en/elections-2023/}}. The tool uses ChatGPT (via its API) to analyze the speeches, identify topics, polarization levels, sentiment, etc \cite{troboukis202}. During the workshop, participants interacted with a demo version of the tool, retaining key visualizations and including question-answer fields\footnote{Demo tool available online: \url{https://case-studies.trustinscience.eu/}}. Demo tool material has been compiled in the Appendix \ref{appendixC}.

\textbf{(A1) Baseline factors and elements collection}. Participants completed three well-established questionnaires about demographics \cite{eurobarometer_2022}, political orientation \cite{smith2013civic}, and attitudes towards AI \cite{schepman2020initial, grassini2023development}. 

\textbf{(A2) Interactive trustee engagement}. Participants engaged in four \textbf{I}nteractions within the demo tool, each carefully crafted for a particular reason and given conditions. \textit{\textbf{I1}-Analysis sufficiency detection}: Participants explored the topics (identified by ChatGPT) mostly referenced by political leaders in their speeches; then participants were asked whether they found the analysis sufficient or preferred to read the speech themselves. The three questions answered (Tab. \ref{tab:workshop_questions}: Q1-Q3) assessed the tool's sufficiency and reported trust-related actions (button-press); \textit{\textbf{I2}-Comparative accuracy assessment}: Participants were evenly divided into control (14) and experimental (15) groups to explore two visualizations of speech polarization. The control group was informed that the results in both visualizations were ChatGPT-generated using different analysis approaches. The experimental group was informed that the results in the first visualization were ChatGPT-generated, while in the second visualization ChatGPT-generated and human-validated. The two questions answered (Tab. \ref{tab:workshop_questions}: Q4-Q5) assessed trust differences based on knowledge of human involvement; \textit{\textbf{I3}-Fully-aware accuracy assessment}: Regardless of their assigned group, participants re-examined the second visualization of \textbf{\textit{I2}}, but in this interaction, all were informed that results were ChatGPT-generated and validated by data journalists. The question answered (Tab. \ref{tab:workshop_questions}: Q6) assessed trust differences based on the knowledge of data journalists involvement; \textit{\textbf{I4}-Opaqueness impact detection}: Participants explored two visualizations showing the sentiment detected in the political speeches of two opposing leaders (governing and opposition party), generated by ChatGPT and validated by data journalists. Unlike previous interactions, they could not hover over the visualizations to read the speeches. The two questions answered (Tab. \ref{tab:workshop_questions}: Q7-Q8) assessed how the lack of oversight to the speeches affected trust. After all interactions, they rated whether they would use the tool again (Tab. \ref{tab:workshop_questions}: Q9).

\textbf{(A3) Trustor-trustee attitudes collection}. Participants completed two well-established questionnaires about their overall trust in the tool \cite{radensky2023think} and their perceptions towards organizations and scientists \cite{funk2019trust}. Finally, they were informed that some of them would be invited for follow-up activities (e.g., interviews).

\subsubsection{Quantitative data analysis} During the workshop, participants' quantitative data were collected via LimeSurvey\footnote{\url{https://www.limesurvey.org/}} and the demo tool logs, then securely stored on university servers and proprietary cloud services. The data were preprocessed for statistical analysis (e.g., data type conversions). For numerical correlations, Spearman’s correlation \cite{spearman1961proof} was applied, as it is well-suited for smaller sample sizes (N=29), non-normal distribution, and preferable ranking reliance \cite{de2016comparing}. For numerical-categorical correlations, point-biserial correlation coefficient \cite{maccallum2002practice} was applied, due to its -1 to 1 output range. For statistical significant differences, confidence intervals were calculated using the t-distribution, given their suitability for smaller sample sizes \cite{de2019using}.

\subsection{Deeper exploration of causal trustor-trustee relationships}\label{interviews}
To gain deeper insights into how trustor (users/participants) contextual factors shape attitudes about trustee (LLMs) systemic elements and their relationships, we conducted online individual semi-structured interviews under the bowtie core. Each interview (\(\sim\)30 minutes) was recorded upon participants' consent. The bowtie model (Fig. \ref{fig:bowtie}) further guided the questions' focus on those contextual factors and systemic elements explored more broadly in the activities (Tab. \ref{tab:interviews_questions}).

\begin{table*}[ht!]
\centering
\scriptsize
\caption{Questions asked and data collected during the interviews, and their alignment with the bowtie model of trust in LLMs.}
\label{tab:interviews_questions}
\begin{tabular}{l|l|l|c|l}
\toprule
\multicolumn{1}{c|}{\textbf{}} &
  \textbf{Question} &
  \multicolumn{1}{c|}{\textbf{Bowtie side}} &
  \textbf{\begin{tabular}[c]{@{}c@{}}1st fold category\end{tabular}} &
  \multicolumn{1}{l}{\textbf{2nd fold factor or element}} \\ \midrule \midrule
\multirow{3}{*}{\textbf{Experiences with AI \& LLMs}} &
  \multirow{2}{*}{\begin{tabular}[c]{@{}l@{}}Familiarity, relationship and views \\ about AI and LLMs\end{tabular}} &
  \multirow{2}{*}{Systemic element} &
  \multirow{2}{*}{Scientific discipline} &
  Attitudes towards LLMs \\ \cline{5-5} & & & &
  Familiarity with LLMs \\ \cline{2-5} &
  Specific experiences with AI and LLMs &
  Contextual factor &
  Background &
  Past experiences \\ \midrule \midrule
\multirow{4}{*}{\textbf{\begin{tabular}[c]{@{}l@{}}Perceptions towards \\ organizations and scientists\end{tabular}}} &
  \multirow{2}{*}{\begin{tabular}[c]{@{}l@{}}Views about AUTh, iMEdD, EU \\ involvement in the study\end{tabular}} &
  Contextual factor &
  \multicolumn{1}{c|}{Perceptions} &
  Funding scheme \\ \cline{3-5} & &
  Systemic element &
  Stakeholders & AUTh, iMEdD, EU \\ \cline{2-5} &
  \multirow{2}{*}{\begin{tabular}[c]{@{}l@{}}Contribution of data scientists, data journalists \\ and political scientists to the tool\end{tabular}} &
  Systemic element & Scientists &
  \begin{tabular}[c]{@{}l@{}}Data scientists, data journalists, \\ and political scientists\end{tabular} \\ \cline{3-5} &&
  Contextual factor & \multicolumn{1}{c|}{Perceptions} &
  Attitudes towards involved scientists \\ \midrule \midrule
\multirow{2}{*}{\textbf{Trust in human-centric elements}} &
  \begin{tabular}[c]{@{}l@{}}Elaborations on accuracy assessments for the \\ HITL-related tasks\end{tabular} &
  Systemic element &
  Scientists &
  \begin{tabular}[c]{@{}l@{}}Data scientists, data journalists, \\ and political scientists\end{tabular} \\ \cline{2-5} &
  \begin{tabular}[c]{@{}l@{}}Elaborations on accuracy assessments in tasks\\ where transparency and oversight are missing\end{tabular} &
  Systemic element &
  Product &
  Political discourse tool \\ \midrule \midrule
\textbf{Recommendations for improvement} &
  \begin{tabular}[c]{@{}l@{}}Overall satisfaction with the tool \\ and weak features\end{tabular} &
  Systemic element &
  Product &
  Political discourse tool \\ \bottomrule
\end{tabular}
\end{table*}

\subsubsection{Interviews procedure} The interviews were conducted via Zoom. They were structured into four parts, each carefully crafted around a theme:
\textbf{Experiences with AI \& LLMs}. Participants shared their views on the increasing popularity of AI and LLMs, and recalled specific experiences with these technologies, enriching information on trustor background factors and enabling an exploration of how past experiences influence trust-related actions and trust in LLMs as a scientific discipline;
\textbf{Perceptions towards organizations \& scientists}. Participants first received detailed information about the contributions of various organizations (AUTh: Aristotle University of Thessaloniki, EU: European Union, iMEdD: Incubator for Media Education and Development) and experts (data scientists, data journalists, political scientists) involved in the tool's development. Then, they shared their views, particularly regarding their competence and credibility, offering deeper insights on trustor perceptions and trust attitudes towards scientists, and stakeholders;
\textbf{Trust in human-centric elements}. After revisiting their responses to \textbf{\textit{I2}} and \textbf{\textit{I3}} (HITL-related interactions), interviewees elaborated on their accuracy assessments to provide deeper insights into the role of human and expert involvement in shaping trust. Similarly, they revisited their responses to \textbf{\textit{I4}}, focusing on the rationale behind their decision-making in the absence of speech and reasoning how human-centric elements influence trust in the tool;
\textbf{Recommendations for improvement}. Participants shared their overall opinions on the political discourse tool, including its use in real-world scenarios (e.g., next elections), enriching information on trust attitudes towards the tool, along with weaknesses and suggestions for improvement.

\subsubsection{Participants and recruitment} We recruited 10 participants for interviews using critical sampling, to explore the cases likely to yield the most important points \cite{patton2014qualitative}, selecting participants whose quantitative data analysis revealed the most critical positioning at the four interview parts. All 10 participants answered questions from all parts. Detailed criteria for criticality and analysis results are provided in Appendix \ref{appendixA}.

\subsubsection{Qualitative data analysis} Recordings were transcribed using \textit{rev} transcription software\footnote{https://www.rev.com/} and manually checked. Identifying information was removed, and transcripts were securely stored on university servers and cloud services. The qualitative data were coded using a deductive thematic analysis \cite{braun2006using}. We established a coding framework rooted in themes defined based on our bowtie model (Fig. \ref{fig:bowtie}): 1) Experiences with AI/LLMs; 2) Actors and stakeholders in the AI/LLM system; 3) AI/LLM human-centric elements. We systematically coded participants' responses to these themes, performing consensus coding to overall enhance the reliability and consistency of the final analysis.

\section{Research Questions: Analysis \& Insights}
Our systematic bowtie model enabled the collection of quantitative and qualitative data to examine: trustor-side contextual factors; trustee-side systemic elements; inter-relationships between the two sides; intra-relationships within each side; and their underlying causal effects. Rigorous correlation and thematic data analytics respond to our RQs (section \ref{conceptualframework}), and provide evidence-based insights highlighted in a series of major takeaways.

\subsection{How do trustor contextual factors influence their trust-related actions? [RQ1]}\label{rq1results}
The particular trustors' contextual factors which influence their decision-making patterns in \textbf{\textit{I1}} (Tab. \ref{tab:workshop_questions}: Q1-Q2) were captured by recording their spontaneous reaction (to either press the button to read the speech or rely on the visualization's analysis) (\textbf{C2}). Among the 29 participants, only two did not press the button to read the speech, explaining that the visualization's analysis was sufficient and aligned with their expectations. Especially, there is a positive predisposition regarding ChatGPT's (LLM embedded in the demo tool) competence to analyze large corpus, with younger trustors stating that it can cope with the \textit{``bulk of information when it comes to an election period''} (P4). Conversely, 27 participants pressed the button to read the speech. Some of them argued being curious about its content and wanting more detailed information beyond the visualization, such as the political party referenced. Others expressed distrust in ChatGPT's analysis, particularly concerning its ability to identify topics accurately and avoid bias, thus reflecting an overall disposition to independently cross-check whether ChatGPT results align with their own judgments. As P6 elaborated: \textit{"If I want to ensure it’s correct, I’ll still verify it. I may ask, "How can I search for this on Google or YouTube?" to get an idea and then evaluate it myself rather than blindly accept what it says. Ok, it’s also helped a lot with coding, but I believe we must understand what we’re doing, not just take the code, have ChatGPT debugging it, and call it a day. After all, ChatGPT, as a neural network, doesn’t truly understand what it’s saying, it predicts the next token based on probabilities."}. Finally, as opposed to simple tasks, there are further negative predispositions and prominent human biases towards ChatGPT's competence in critical tasks such as political discourse analysis. These are seen as better suited for humans (\textit{"for critical thinking, I think a human mind should do the job, you can't just provide answers from a machine",} P8). \textbf{\underline{Takeaway \#1}: Trustors negatively inclined to trustee's competence, results validity, and alignment with their judgments, expressed their curiosity for further interaction and justifications by the trustee before engaging.}

\subsection{What is the mapping between trustor contextual factors and trustee:}\label{rq2results}
\subsubsection{\textbf{systemic elements? [RQ2A]}}\leavevmode \newline \label{rq2Aresults}
Correlations between participants' contextual factors and LLMs' systemic elements uncover the trustor-trustee bidirectional inter-relationship (\textbf{C2}), revealing that \textbf{not all contextual factors influence equally the different systemic elements}. 

\begin{figure*}[ht!]
    \centering
    \subfloat[Scientific discipline-AI/LLMs]{\label{sfig:science}\includegraphics[width=.49\textwidth]{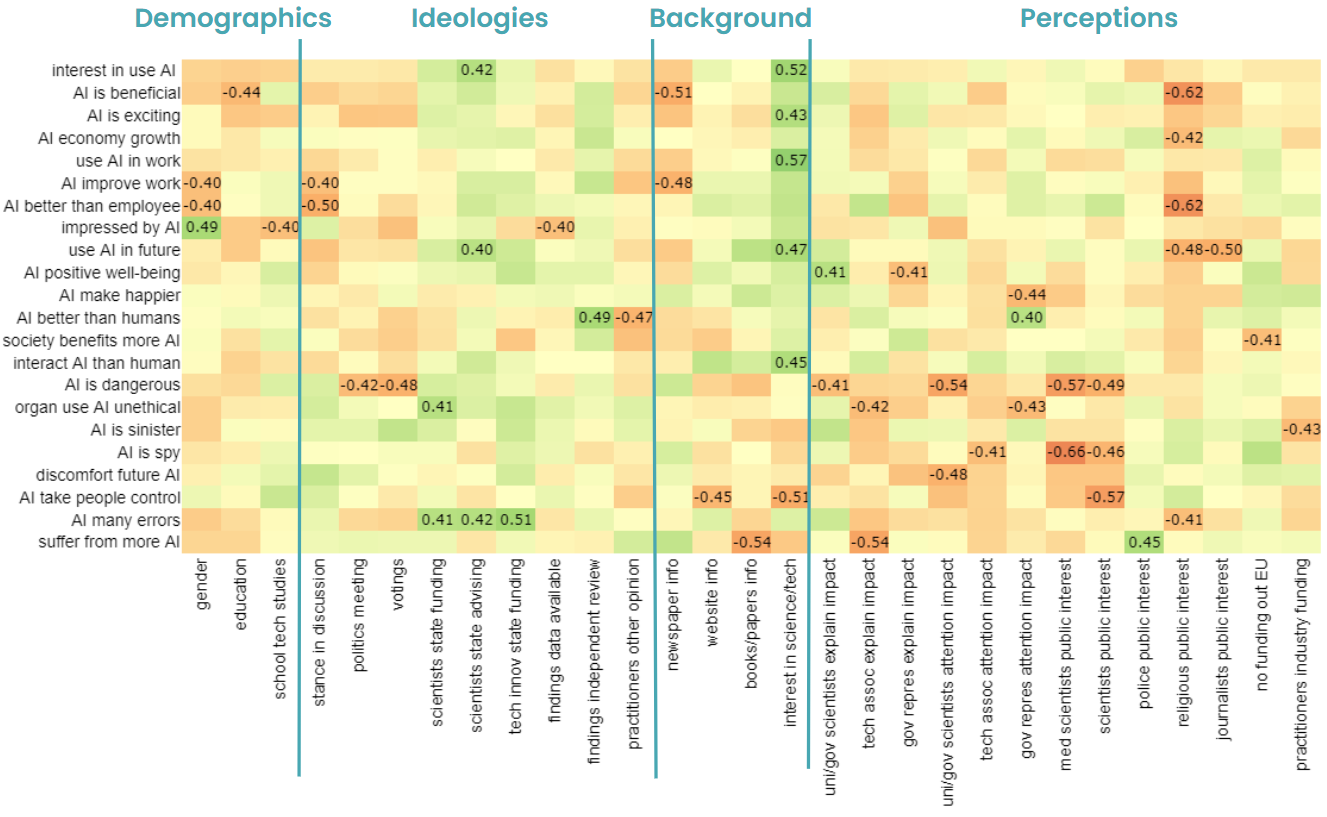}}\hfill
    \subfloat[Products-Political discourse tool]{\label{sfig:tool}\includegraphics[width=.49\textwidth, height=5.3cm]{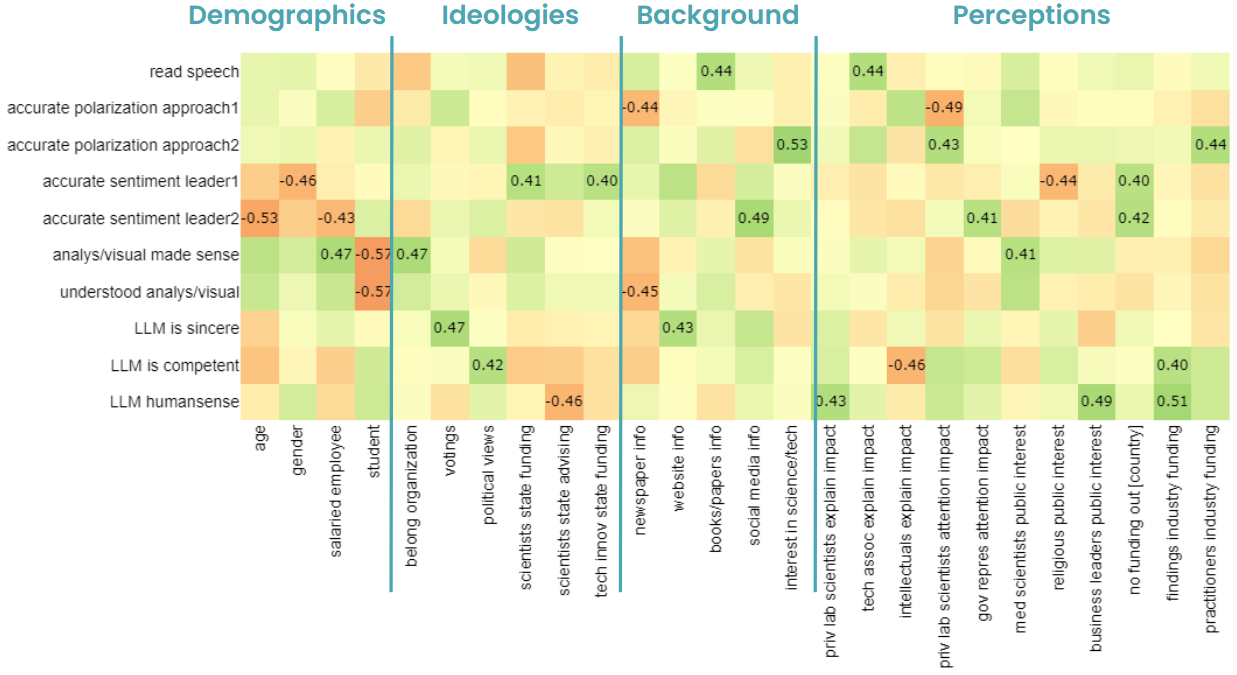}}\\
    \caption{The moderate and high correlations of contextual factors (per category) with the two systemic elements with quantitative data: scientific discipline (AI/LLMs) \ref{sfig:science} and products (political discourse tool) \ref{sfig:tool}.}
    \Description{The figure consists of two heatmaps showing moderate and high correlations between contextual factors and two systemic elements. The left heatmap represents correlations with the scientific discipline (AI/LLMs), while the right heatmap represents correlations with the political discourse tool. On the y-axis are the variables of each element and on the x-axis are the contextual factors organized in their respective contextual categories: demographics, ideologies, background, and perceptions. Correlation values are shown in color-coded intensity on the heatmaps.}
    \label{fig:rq2a}
\end{figure*}

Figure \ref{fig:rq2a} highlights moderate and strong \cite{schober2018correlation} statistical correlations among all contextual factors and the elements of scientific discipline (AI/LLMs) (Fig. \ref{sfig:science}) and products (Fig. \ref{sfig:tool}). The rest are qualitatively analyzed. As expected, trustors having studied science/technology at school are less impressed by AI (Fig. \ref{sfig:science}: r=-0.40) and have more realistic expectations about LLM capabilities. Qualitative insights additionally reflect that those expanding such studies at university further acknowledge that LLMs are often \textit{``faulty''} (P1) and their performance strongly relies on how users will prompt them and divide a complex task into simple subtasks. They also comprehend in depth how LLMs operate, thus do not perceive them as \textit{``god-like tools magically giving perfect answers''} (P6), but stress this may not hold for people lacking technical knowledge: \textit{``When I first entered this field, I found it strange how it could produce coherent answers. Once I started understanding how it works and demystified it, I thought, ``Okay, it’s impressive, but it’s not something alien-like.'' But unfortunately, for most people, for someone not involved in the field, it will likely remain a black box because it’s not easy to learn about, nor can it be explained easily. That’s where questions and misunderstandings often arise.''} (P6). Also, the trustors who are interested in science/technology, are more interested (Fig. \ref{sfig:science}: r=0.52) and excited (Fig. \ref{sfig:science}: r=0.43) are about AI, declaring that they would use AI in their jobs (Fig. \ref{sfig:science}: r=0.57), and generally in the future (Fig. \ref{sfig:science}: r=0.47). They also expressed satisfaction, even noting feeling more \textit{``productive''} (P1) and that \textit{``they make life easier in day-to-day tasks''} (P4). While they do not believe that AI will take control of people (r=-0.51), they stressed the prevalence of misinformation and misconceptions around AI (\textit{``most people nowadays are afraid that AI will take over the world'',} P2), noting that such fears are exaggerated and often stem from a lack of awareness of how AI operates. Finally, trustors who express greater trust in university or government scientists and technology associations exhibit more positive views about AI and LLMs, as they are less likely to believe that AI: is dangerous (Fig. \ref{sfig:science}: r=-0.54), spies on people (Fig. \ref{sfig:science}: r=-0.46), causes people to suffer (Fig. \ref{sfig:science}: r=-0.54), takes control of people (Fig. \ref{sfig:science}: r=-0.46),evokes discomfort in the future applications (Fig. \ref{sfig:science}: r=-0.48). \textbf{\underline{Takeaway \#2}: Trust in LLMs scientific discipline is shaped by trustors' expertise, interest in LLMs, and positive perceptions of LLMs scientists.}

Interestingly, trustors more engaged in politics exhibit greater trust in the LLM-based political discourse tool; its perceived competence increases and its results made much sense (both in Fig. \ref{sfig:tool}: r=0.47). Qualitative data also revealed that those having an informed opinion on the political scene of Greece interpret and cross-check the tool's results easier. Additionally, trustors interested in science/technology found the second approach of polarization analysis (\textbf{\textit{I2}} - Tab. \ref{tab:workshop_questions}: Q5) more accurate (Fig. \ref{sfig:tool}: r=0.53) - potentially influenced by their positive views about scientists' competence in general (Fig. \ref{sfig:tool}: r=0.43) and thus their involvement in this analysis. \textbf{\underline{Takeaway \#3}: Trust in the tool is mostly affected by trustor domain knowledge (politics) and their perceptions about scientists' involvement, rather than the tool's technological competence.}

Then, trustors' attitudes varied across different categories of scientists contributing to the tool. Data scientists received moderate attention, eliciting discussion around the tool's features, satisfaction with its design (\textit{``It's very clever,''} P6), and refinements for utility optimization. Data journalists received greater attention with mixed reactions; their expertise and involvement were acknowledged, but concerns arose about potential political biases and their ability to remain \textit{``free of speech''} (P5). Comparably, political scientists were perceived more positively. Their expertise was seen as more relevant to the tool, and no concerns existed about their capacity to deliver unbiased analyses \textit{(``political scientists can be more clear on what they are studying and reviewing. I mean, how else could you make it better?''}, P2). \textbf{\underline{Takeaway \#4}: Trust in scientists varies: no concerns were expressed for data scientists, major concerns existed for data journalists' political biases, and highly positive statements were made for political scientists' involvement.}

Finally, trustors' attitudes about the different stakeholders of LLMs (and our LLM-based tool) are also mixed. Skepticism exists about private technology companies - as opposed to public entities - conducting research in AI and LLMs due to potential conflicts of interests \textit{(``profit is a big factor, so some companies may not take the best approach for the general good''}, P3). Also, while the industry fosters innovation, there are \textit{``concerns about the free use, exploitation, of such outcomes''} (P10). Trustors also note some prerequisites for LLMs stakeholders, including \textit{``setting right boundaries, especially in high-risk domains''} (P2), and investing in \textit{``humans-AI collaboration''} (P10). \textbf{\underline{Takeaway \#5}: Trust in LLMs' stakeholders variates with respect to trustors' views about stakeholders' ethical norms and motivations.}

\subsubsection{\textbf{human-centric systemic elements? [RQ2B]}}\leavevmode \newline \label{rq2Bresults}
Exploring how trustors perceive the tool's HITL conditions and lack of transparency based on their contextual factors, we uncover insights into the trustor-trustee inter-relationship (\textbf{C2}), and deliver evidence-based insights for practitioners and policymakers (\textbf{C3}).

\begin{figure}[ht!]
    \centering
    \includegraphics[width=0.4\textwidth, height=4cm]{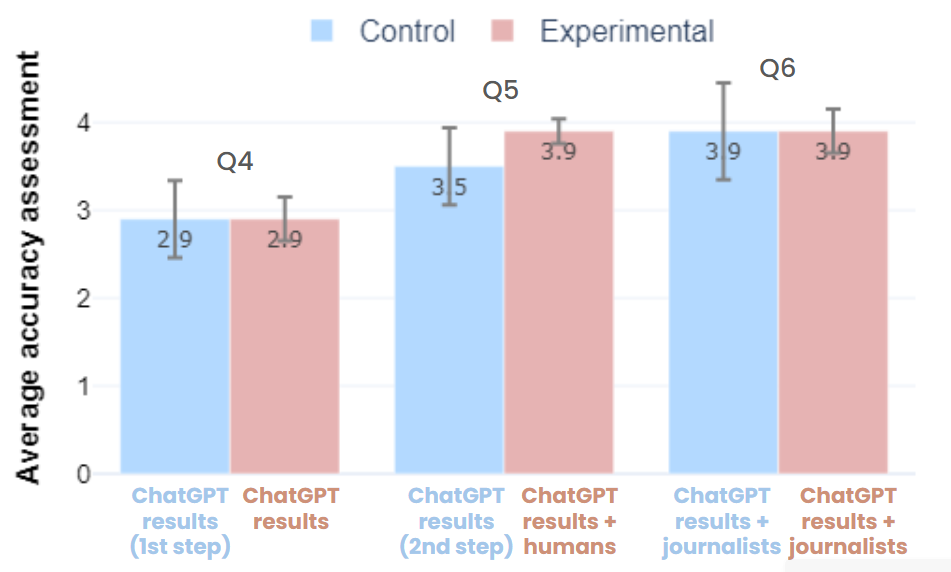}
    \caption{Avg. accuracy assessment for control and experimental groups, with conf. int., in different HITL conditions.}
    \Description{A bar chart comparing average accuracy assessments for control and experimental groups across three questions: Q4, Q5, and Q6. The x-axis shows the available information for each group. In Q4, for control: ChatGPT results (1st step), for experimental: ChatGPT results; in Q5, for control: ChatGPT results (2nd step), for experimental: ChatGPT results + humans; in Q6, for both control and experimental: ChatGPT results + journalists. The y-axis represents average accuracy scores. Both groups assessed Q4 at 2.9, Q5 with control at 3.5 and experimental at 3.9, and Q6 equally at 3.9.}
    \label{fig:human-in-the-loop}
\end{figure}

\textbf{Human-in-the-loop (HITL).} 
To explore the influence of the HITL approach on trustors' attitudes, we analyze the average accuracy assessments from \textbf{\textit{I2}} (Tab. \ref{tab:workshop_questions}: Q4-Q5) and \textbf{\textit{I3}} (Tab. \ref{tab:workshop_questions}: Q6), where participants were provided with varying conditions of information on humans' and data journalists' involvement (details in \ref{workshop}), as presented in Fig. \ref{fig:human-in-the-loop} and Tab. \ref{tab:human-in-the-loop}.

\begin{table}[ht!]
    \centering
    \scriptsize
    \caption{Average accuracy, confidence intervals (CI), and range (min and max values) for control and experimental groups across different HITL conditions (questions).}
    \label{tab:human-in-the-loop}
    \begin{tabular}{r|c|c|c|c|c}
    \textbf{Group} & \textbf{Question} & \textbf{Avg. acc.} & \textbf{CI} & \textbf{min} & \textbf{max} \\ \toprule
    \multirow{3}{*}{Control} & Q4 & 2.86 & 0.44 & 2.42 & 3.30 \\
                             & Q5  & 3.5  & 0.44 & 3.06  & 3.94  \\
                            & Q6  & 3.86 & 0.55  & 3.31  & 4.41  \\ \midrule \midrule
    \multirow{3}{*}{Experimental} & Q4  & 2.93  & 0.25  & 2.68  & 3.18  \\
                                  & Q5  & 3.93 & 0.14 & 3.79  & 4.07 \\
                                  & Q6  & 3.93 & 0.25 & 3.68 & 4.18        
    \end{tabular}
\end{table}

The control group found the results more accurate when informed about data journalists' involvement, compared to results produced in the first approach of ChatGPT analysis. As P10 explains, experts \textit{``have the background and knowledge to evaluate this data''}. The experimental group clearly assessed ChatGPT-generated results as more accurate when validated by either humans (opposite to control) or data journalists (similar to control). However, the experimental group showed no accuracy assessment increase when informed that the validators were data journalists, likely due to concerns about human objectivity in analyzing political discourse: \textit{``People cannot judge fully objectively. We all have an opinion, even political scientists, journalists, everyone has one, and it’s difficult to be fully objective on things like that, while LLMs, machines in general, can consider things more objectively than us''} (P5). As previously argued (see \ref{rq2Aresults})), especially data journalists trigger skepticism about their ability to provide unbiased analyses free of political orientations. As P4 remarked, \textit{``when you are a person that has to review something like this and hear the word journalist, sometimes you're biased''}. Further details on how trustors’ particular contextual factors influence perceptions of the HITL approach are provided in Appendix \ref{appendixB}. \textbf{\underline{Takeaway \#6}: Trust significantly increases with the HITL approach embedded in the LLM-based tool, with human involvement having a subtly more positive impact than experts' involvement.}

\textbf{Transparency.}
To examine how the lack of transparency affects trustors' perceived accuracy, we compare the average accuracy assessments from \textbf{\textit{I3}} (Tab. \ref{tab:workshop_questions}: Q6) and \textbf{\textit{I4}} (Tab. \ref{tab:workshop_questions}: Q7-Q8), where all participants were informed that results were ChatGPT-generated and validated by data journalists from iMEdD (details in \ref{workshop}). However, the two visualizations in \textbf{\textit{I4}} lacked the option to view the speech by hovering over, limiting user oversight and evoking a consistent decline in accuracy assessments, from 3.9 to 3.76 and further to 3.17 in Q6-Q8 respectively. Lacking details about how and why the speeches were categorized as such, most trustors relied on their own background knowledge of the rhetorical strategies typically employed by the two politicians. Thus, a few participants expected the governing party's speech (\textbf{\textit{I4}}: left) to be categorized as more positive, discouraging them from assigning a high accuracy score. Similarly, some questioned the neutrality of political speeches -especially from the opposition party (\textbf{\textit{I4}}: right)- , noting their reliance on rhetorical techniques over factual reporting. Participants further stated that opposition party's speeches should be reasonably characterized as more negative, as they deliver strong criticism of the party in power (\textit{``He really tries to persuade people that everything that the governments is acting on right now, it's wrong. So, I don't think it's neutral.'', P1}). Some trustors were troubled by the absence of transparency and assigned the same scores to both visualizations, avoiding more detailed judgments. As P10 explained, they were confident in the analysis produced through \textit{``AI-human experts collaboration''} and assigned equally high scores. Overall, most trustors would prefer to see the speeches informing the visualizations to provide more informed assessments, with P4 suggesting that \textit{``if there was text, I would trust my answer more''}. \textbf{\underline{Takeaway \#7}: In response to the lack of transparency, trustors rely on their knowledge of Greece's political affairs to assess results accuracy.}

\subsection{How does trustor trust attitudes differentiate over the trustee systemic elements? [RQ3]}\label{rq3results}
\begin{figure}[ht!]
    \centering
    \includegraphics[width=0.4\textwidth, height=4cm]{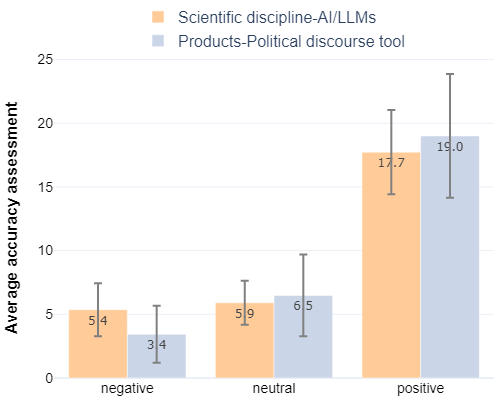}
    \caption{Avg. accuracy assessment per category (negative, neutral, positive) for the two systemic elements: scientific discipline and products, with their confidence intervals.}
    \Description{The figure presents a grouped bar chart showing the average accuracy assessments across three categories (negative, neutral, positive) for two systemic elements: scientific discipline (AI/LLMs) and products (political discourse tool). The x-axis represents the three categories, the y-axis shows average accuracy assessment values and confidence intervals are displayed as error bars. For both elements the positive assessments are higher.}
    \label{fig:rq3_all}
\end{figure}

Examining trustors’ attitudes toward the different trustee-side systemic elements, we uncover the intra-relationship within the trustee (LLMs) (\textbf{C2}). Quantitative Likert-scale data collected (Tab. \ref{tab:workshop_questions}) for scientific discipline (AI/LLMs) and products (political discourse tool) are categorized into negative, neutral, and positive. Then, we compare trustors’ attitudes towards them (Fig. \ref{fig:rq3_all}). The elements of scientists and stakeholders are qualitatively analyzed.

\begin{figure}[ht!]
    \begin{subfigure}{0.4\textwidth}
        \includegraphics[width=0.99\textwidth, height=3cm]{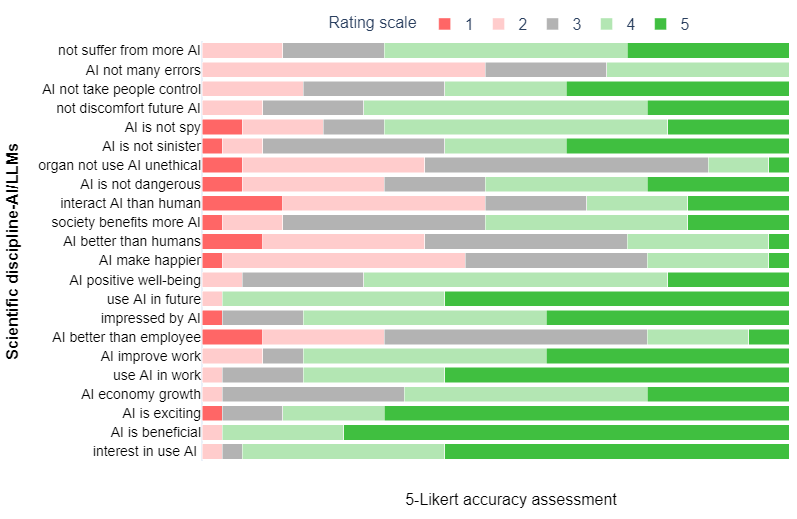}
        \caption{Scientific discipline-AI/LLMs}
    \end{subfigure}
    \hfill
    \begin{subfigure}{0.4\textwidth}
        \includegraphics[width=0.99\textwidth, height=2cm]{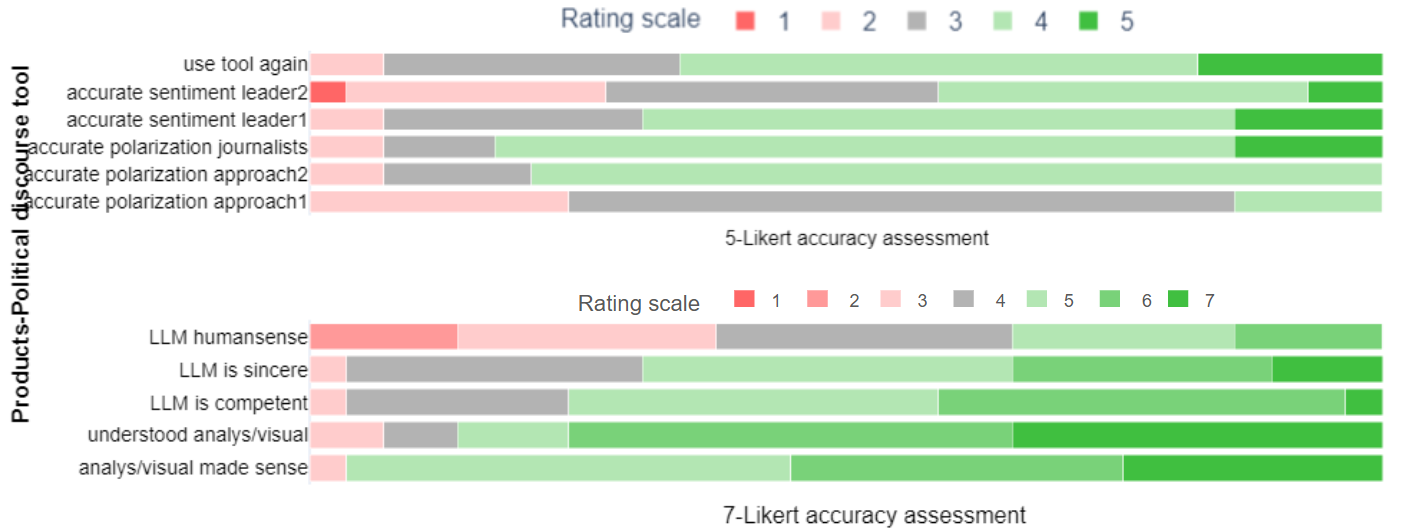}
        \caption{Products-Political discourse tool}
    \end{subfigure}
    \caption{The distribution of participants answers to each specific item of the two systemic elements: scientific discipline-AI/LLMs and products-political discourse tool.}
    \label{fig:likerts}
    \Description{The figure displays two sets of Likert scale assessments. The first part (a) shows the distribution of participant answers for the ``Scientific discipline-AI/LLMs'' category, with a 5-point Likert scale, ranging from ``1'' to ``5'', for various items such as ``AI is exciting'', ``AI is beneficial'', and ``interest in use AI''. The second part (b) shows the distribution of answers for the ``Products-Political discourse tool'' category, with a 7-point Likert scale, ranging from ``1'' to ``7'', for items like ``LLM is competent'', ``use tool again'', and ``accurate sentiment leader''. Both sections present color-coded bars representing the response distribution for each item.}
\end{figure}

Regarding the LLMs scientific discipline and LLM-based political discourse tool, the positive attitudes are more statistical significant than the neutral and negative ones, indicating general trust in LLMs. Only with marginal (but not statistical significant) differences, trust in the tool seems higher than trust in the LLMs discipline. Thus, we employ qualitative insights to substantiate these differences. LLMs as a scientific discipline is mostly viewed positively by trustors due to LLMs' beneficial contribution (e.g., in work-related tasks (relevant Fig. \ref{fig:likerts}a rows)), with P4 arguing \textit{``they make my life easier in day-to-day tasks''} and P6 stating: \textit{``Of course we lived without it before, but it’s here to make life easier to some extent, not to take away jobs or harm us, as some claim, which often makes me say, ``Whoa!''}. 

Similarly, the tool is perceived positively. Most participants would use it again, find its results accurate and well-presented (Fig. \ref{fig:likerts}b), and praise its ability to summarize political information and help formulate informed opinions. Some even claimed it can perform such analyses more objectively than humans (\textit{``It's the most objective view on the political matters we can get right now''}, P5) - although this may also reflect over-reliance on LLMs' capabilities and a not so well-grounded trust attitude toward the tool. Interestingly, trustors' views about the tool are unconsciously influenced by their attitudes towards the LLMs scientific discipline. For instance, they note the impossibility of fully trusting such tools (\textit{``I don't think I will give a five never because it's a piece of software''}, P1), and are hesitant to declare complete trust in the tool's results, as this means they totally align with their personal beliefs (\textit{``I can't change my rating to five because this thing suggests that I definitely believe the results are the same as my belief''}, P3). This evidence reveals that while trustors consciously assess their trust to an LLM-based tool based on its particular features (visualizations, accuracy), they are unconsciously affected by more generic attitudes towards LLMs broader competence and alignment with their own judgments.

Finally, the element of scientists is generally positively valued, as RQ2A and RQ2B (see \ref{rq2Aresults} \& \ref{rq2Bresults}) also reveal. As further evidenced below, trustors' perceptions about scientists critically shape their attitudes towards specific stakeholder elements and corresponding organizations, due to their inherent overlaps. For instance, positive trust attitudes about scientists are largely intertwined with trustors' views about public universities. As P1 noted, \textit{``public universities are trying to create an impact on the field and there are people really trying to make a difference''}. Public universities also remain the most positively regarded compared to other stakeholders of our model, described as ``\textit{the best players in the game as they can assure we respect all human rights''} (P5). Then, trustors' concerns about data journalists extend to journalistic organizations, similarly reflecting how these two elements are intertwined. Finally, private entities within the technology industry trigger skepticism -as argued in RQ2A (see \ref{rq2Aresults}), and European funding agencies surprisingly elicit controversial perspectives. They are viewed as having both a leading and restrictive contribution in LLM research, with P1 explaining: \textit{``European Union has a very supportive role, but I don't think it really helps innovation and research in the field. I think a group of people that are trying to regulate and restrict every research that comes from the other side of the sea, of the Atlantic, they're trying to put boundaries in everything that comes from there, and aren't really trying to create solutions''} (P1). \textbf{\underline{Takeaway \#8}: Trustee systemic elements evoke differentiated trust attitudes, and exhibit overlaps in how they are perceived due to their inherent connections.}

\section{Discussion \& Future Work}\label{discussion}
We now discuss key insights from our findings, offering practical recommendations to researchers, practitioners, policymakers, and the general public (\textbf{\textit{C3}}), followed by limitations and future work.

\subsection{Impossibility of universal trust in LLMs}\label{discussion2}
Our exploration of the complex sociotechnical topic of trust in LLMs, in the high-stake political discourse, has yielded a fine-grained understanding of trustor-trustee roles and their relationships. As evidenced, trust manifests as a multi-faceted and intricate form, which we could not holistically explore without the proposed bowtie model, since its core is needed to robustly tie the trustor-trustee sides and uncover their hidden relationships. Concurrently, these relationships clearly revealed that achieving complete and universal trust in LLMs is an almost impossible and even utopic endeavour, since both trustor-side contextual factors and trustee-side systemic elements have strong inherent connections. Our findings suggest that trustors' expertise while initially enhances LLMs' comprehension and fosters grounded trust attitudes, it may further lead them to rely on their own reasoning, limiting higher trust. This aligns with the inverted-U shape theory \cite{yerkes1908relation}, as applied to trust in LLMs, where greater knowledge initially increases trust but eventually decreases it beyond a certain threshold. Impossibility of universal trust is also evidenced by the congenital nature of LLMs, prohibiting the sharp distinction of each element's role within this complex sociotechnical topic. Namely, LLMs challenge traditional interpretations of agency \cite{suarez2019tay}; their opacity, and lack of intentionality -in principle a human capability - make it difficult to trace their behaviour, root causes and responsible stakeholders - ultimately hindering trustors' ability to form fully informed opinions.

\subsection{Applicability of the bowtie model}\label{applicability}
Our bowtie model's demonstrated applicability and extensibility provide a solid foundation for reproducibility and analytical generalizability, positioning it as a prominent baseline for future trust-related research across various use cases and domains. The model's systematic identification of factors and elements, progressive unfolding, and hierarchical positioning offer an adaptive, dynamic approach to evolving sociotechnical trust topics, within and beyond LLMs. For example, our bowtie model can effectively explore “trust in LLM4Science” within biomedical contexts, positioning patients as the trustor and LLM-driven biomedicine as the trustee. Contextual factors could expand to encompass patient medical history, while ideological components could reflect patient views on medical ethics and reliability. Similarly, systemic elements, such as scientific discipline and products, could further consider LLMs and bioinformatics as separate influential domains and examine LLM-driven tool’s performance, fairness and bias issues, and human-computer interaction elements, respectively.

\subsection{Evidence-based implications}\label{practicalimplications}
Our key takeaways on trustor-trustee relationships highlight the bowtie-based systematic and evidence-based knowledge and insights revealed, targeting diverse stakeholders needs and impact on trust in LLMs. Firstly, \textbf{\textit{policymakers}} with their targeted policy interventions, play a crucial role in shaping trust in LLMs, since our findings underscore the significant impact of trustor contextual factors on trustee-side perceptions. Strengthening public AI literacy -particularly among non-experts- is a key intervention, given the prevalence of deeply-rooted dispositions towards AI and LLMs (takeaway \#1), and the risk of misconceptions among those without technical expertise (takeaway \#2). For instance, both formal and informal educational initiatives on AI and LLMs could play a crucial role in fostering informed engagement and mitigating misunderstandings. Secondly, our findings strongly suggest that LLM \textbf{\textit{practitioners}} should incorporate HITL approaches in LLM-driven tools, as these significantly enhance trustors' trust attitudes (takeaway \#6). Furthermore, they should adopt transparent methodologies to mitigate reliance on individual understanding or domain-specific knowledge, thus reducing human biases (takeaway \#7). Integrating these human-centric practices allow organizations using/developing AI- and LLM-driven tools to align with emerging global regulatory standards on trustworthy AI (e.g., AI Act), highlighting the practical and legal relevance of insights derived from our bowtie model. Thirdly, \textbf{\textit{general public}} awareness of trustor contextual factors and trustee systemic elements is of paramount importance, as our findings indicate that public trust is shaped by understanding and predispositions (incl. biases), particularly in high-stake political discourse (takeaway \#3). Ongoing engagement and participation in discussions can also help clarify controversial topics and perspectives, fostering a stronger trust relationship between the public and diverse stakeholder groups (takeaway \#5). Enhancing the public's understanding of LLMs, their potential concerns, and the current domain-specific landscape could contribute to more transparent trustor-trustee relationships, reducing skepticism and encouraging balanced, critical views on trust in LLMs.

\subsection{Limitations \& Future work}\label{futurework}
Our bowtie model and subsequent study provide valuable insights, yet several improvements are possible. First, while our mixed-method study design achieves considerable analytical generalizability \cite{onwuegbuzie2009call, creswell2007mixed} -where findings support the theoretical and modeling assumptions about the studied phenomenon and their applicability to other contexts- scaling up the sample size can further enhance statistical generalizability. Future studies could extend to larger, diverse, and longitudinal participant groups and data samples. Second, we collected quantitative data for two systemic elements (scientific discipline and products) as a proof-of-concept, which can be further extended to enable further quantitative correlations and comparisons. The choice made was driven by the highly granular nature of each systemic element and by the inherent interconnections in sociotechnical topics. Future research should further explore whether specific and generalizable distinctions between trustee-side systemic elements can be achieved, particularly in critical, dynamic sociotechnical environments, as well as how experimental design factors (e.g., task complexity) can influence trust assessments.

\section{Conclusion}\label{conclusion}
Ties of trust in LLMs evolve rapidly, demanding new approaches able to reveal the limitations and potential of trust attitudes on LLMs. In this work we introduced a systematic bowtie model to holistically examine the complex sociotechnical topic of trust in LLMs, tying trustor-side contextual factors and trustee-side systemic elements. Through the bowtie core -a sequential mixed-methods explanatory study- we uncovered the critical trustor-trustee intra- and inter-relationships. We revealed that: trustors actions are primarily influenced by their predisposition towards trustee's competence; not all trustor contextual factors influence equally the different trustee systemic elements; the HITL approach embedded in our LLM-based tool significantly increases trustors trust attitudes, while lack of transparency decreases them; and trustee systemic elements face inherent overlaps, prohibiting their clear distinction in complex sociotechnical topics. We also delivered evidence-based insights to researchers, practitioners, and policymakers to encourage further development and research in the field.

\begin{acks}
Funded by the Inspiring and Anchoring Trust in Science (IANUS, 101058158) grant, funded by the European Union. Views and opinions expressed are however those of the author(s) only and do not necessarily reflect those of the European Union or the European Research Executive Agency (REA). Neither the European Union nor the granting authority can be held responsible for them.
\end{acks}

\bibliographystyle{ACM-Reference-Format}
\bibliography{bibliography}

\appendix

\section{Recruitment of interview participants}\label{appendixA}
For each interview part we analyzed participants whose quantitative data from the respective activities and interactions to reveal the most critical positioning, as follows:
\begin{itemize}[topsep=5pt]
    \item \textbf{Experiences with AI \& LLMs}. From \textbf{(A2)} and specifically \textit{\textbf{I1}}, critical cases were two participants who did not click the button to read the speech (both invited, one interviewed).
    \item \textbf{Perceptions towards organizations \& scientists}. From \textbf{(A3)} and specifically the questionnaire for trust in organizations, critical cases were three participants expressed high confidence in journalists (three invited, two interviewed), and eight participants rated scientists in private labs as having low qualifications in explaining impact on society (four randomly selected interviewed).
    \item \textbf{Trust in human-centric elements}. From \textbf{(A2)} and specifically \textit{\textbf{I2}} and \textit{\textbf{I3}}, critical cases included:
    \begin{itemize}
        \item two participants who assessed the accuracy of data journalist-validated results lower than ChatGPT-only results (both interviewed);
        \item ten participants in the control group who assessed the accuracy of second-approach results higher then first-approach results (two randomly selected interviewed);
        \item one experimental group participant who assessed the accuracy of human-validated results lower than ChatGPT-only;
        \item two experimental group participants who assessed the accuracy of data journalist-validated results lower that human-validated (both invited, one interviewed).
    \end{itemize}
   From the \textit{\textbf{I4}} critical cases are nine participants who assessed the accuracy of both visualizations lower than in prior interactions -we assume due to the absence of hover-text speech- (three randomly selected interviewed).
    \item \textbf{Recommendations for improvement}. From \textbf{(A2)} and specifically the question about reusing the tool (Q9), critical cases were two participants who gave a low score (2/5) (both invited, one interviewed).
\end{itemize}

\section{Trustor factors and the HITL approach}\label{appendixB}
\begin{figure*}[ht!]
    \centering
    \subfloat[Demographics]{\label{sfig:demographics}\includegraphics[width=.39\textwidth, height=4.5cm]{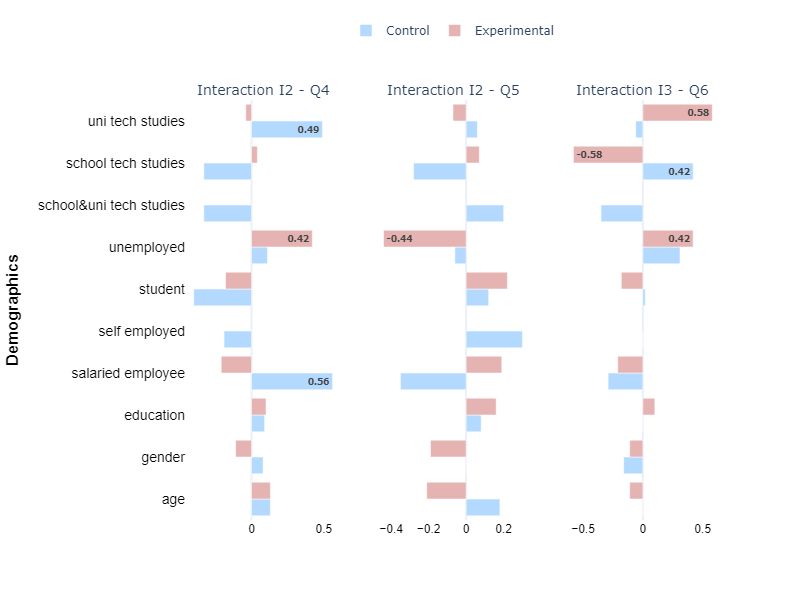}}\hfill
    \subfloat[Background]{\label{sfig:background}\includegraphics[width=.39\textwidth, height=4.5cm]{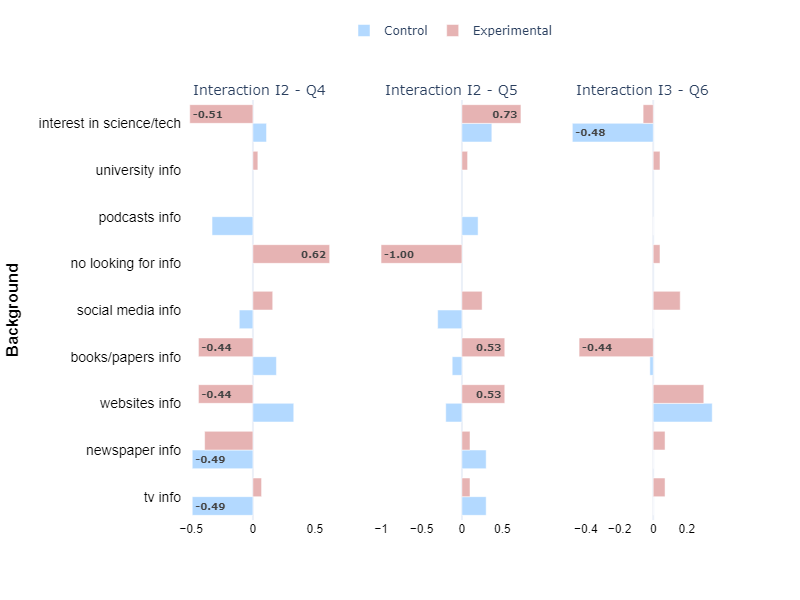}}\\
    \subfloat[Ideologies]{\label{sfig:ideologies}\includegraphics[width=.49\textwidth, height=9cm]{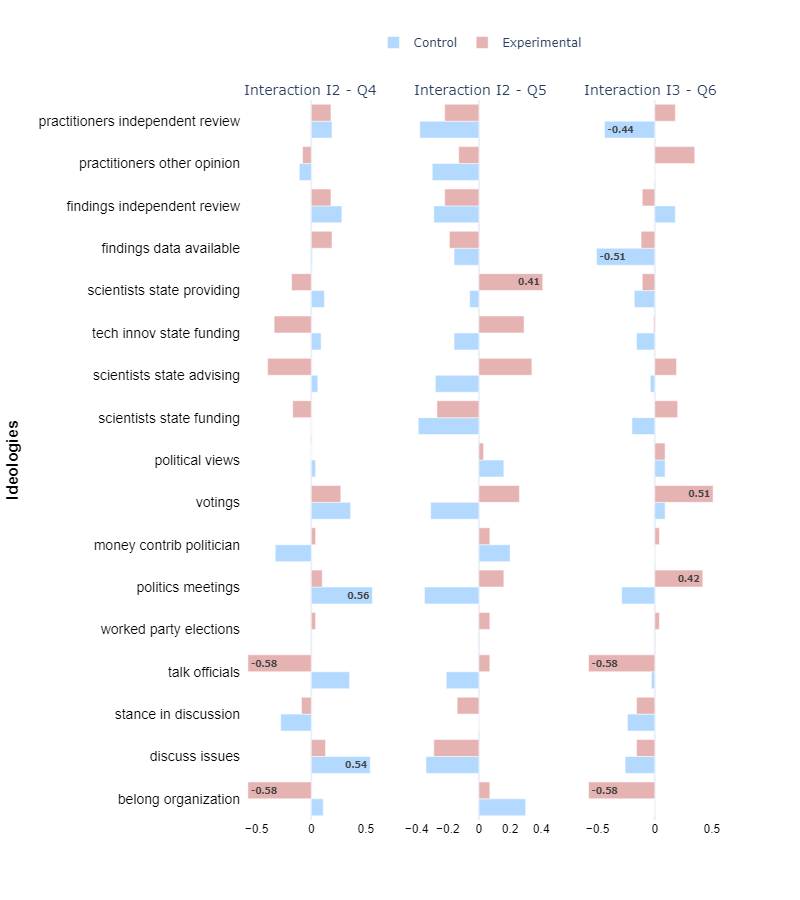}}\hfill
    \subfloat[Perceptions]{\label{sfig:perceptions}\includegraphics[width=.49\textwidth, height=9cm]{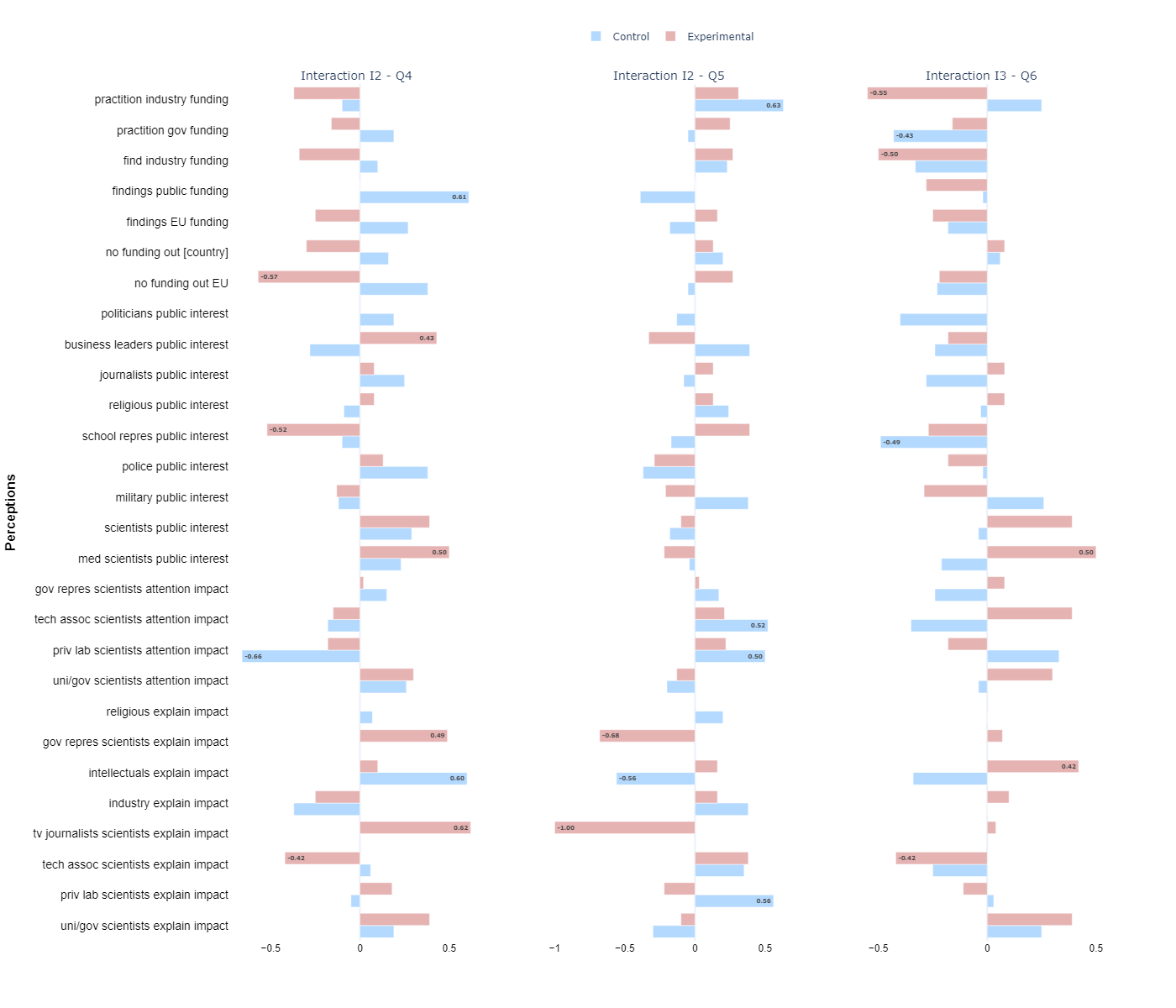}}\\
    \caption{The moderate and high correlations of HITL-related questions with the four categories of contextual factors: demographics (\ref{sfig:demographics}), ideologies (\ref{sfig:ideologies}), background (\ref{sfig:background}), and perceptions (\ref{sfig:perceptions}).}
    \Description{The figure consists of four grouped horizontal bar charts presenting the moderate and high correlations of HITL-related questions (Q4-Q6) with the four categories of contextual factors: demographics (a), ideologies (b), background (c), and perceptions (d). Each chart compares the control group (blue bars) and the experimental group (red bars).}
    \label{fig:contextualfactors_hitl}
\end{figure*}

Trustors' attitudes towards the HITL approach further vary based on their specific contextual factors. Figure \ref{fig:contextualfactors_hitl} highlights the moderate and strong \cite{schober2018correlation} correlations among all contextual factors and the three HITL-related questions (Tab. \ref{tab:workshop_questions}: Q4-Q6). Firstly, regarding demographic factors, trustors who studied science/technology at the university find the results generated by ChatGPT and validated by data journalists (\textbf{\textit{I3}} - Tab. \ref{tab:workshop_questions}: Q6) more accurate (Fig. \ref{sfig:demographics}: r=0.58), whereas those with science/technology education limited to school found these results less accuare (Fig. \ref{sfig:demographics}: r=-0.58). Secondly and most importantly, regarding background factors, trustors closer to science/technology are not in favor of data journalists involvement (\textbf{\textit{I3}} - Tab. \ref{tab:workshop_questions}: Q6), since those more interested in science/technology (Fig. \ref{sfig:background}: r=-0.48) and sourcing information from scientific books and papers (Fig. \ref{sfig:background}: r=-0.44) assessed the accuracy of these results lower. As P1 explained, ChatGPT can execute such an analysis effectively - being a \textit{``simple''} task - with no particular support from human intervention (\textit{``The journalists don't need to add something there because ChatGPT has already done it for them''}). Moreover, P4 relied on his own interpretation of polarization evident in the speech, also because no detailed information was provided on the data journalists' methodology to validate the results: \textit{``It was eyeopening to me that I learned that it was validated by data journalists. But I thought that it was more important to answer as if that was my reaction to seeing those results in terms of polarization. And I was not completely aware of how data journalists produced those results. It was not very clear to me the processes, the methodology they could use. So, it was just a name on top of that plot. It was not much more to guide me to trust one or the other. So, I trusted my initial instinct.''} Still, most participants assessed the accuracy of the results when validated by humans (Fig. \ref{sfig:background}: r=0.73 and r=0.53, respectively) higher when compared to ChatGPT-only results (Fig. \ref{sfig:background}: r=-0.51 and r=-0.44, respectively). Thirdly, ideology-related factors, particularly political engagement, significantly influence perception of HITL approach; participants voting more frequently (Fig. \ref{sfig:ideologies}: r=0.51) and actively participating in political meetings (Fig. \ref{sfig:demographics}: r=0.42) find the results generated by ChatGPT and validated by data journalists (\textbf{\textit{I3}} - Tab. \ref{tab:workshop_questions}: Q6) more accurate. Finally, in terms of perception factors, participants who trust more scientists in private labs and technology associations to explain or address societal impacts rated the second approach of analysis more favorably (Fig. \ref{sfig:perceptions}: r=0.56, r=0.52, and r=0.50, respectively).

\section{Demo tool material}\label{appendixC}
\begin{figure*}[ht!]
    \centering
    \subfloat[\textbf{I1}-Topic analysis sufficiency detection]{\label{sfig:task_1}\includegraphics[width=.49\textwidth, height=6cm]{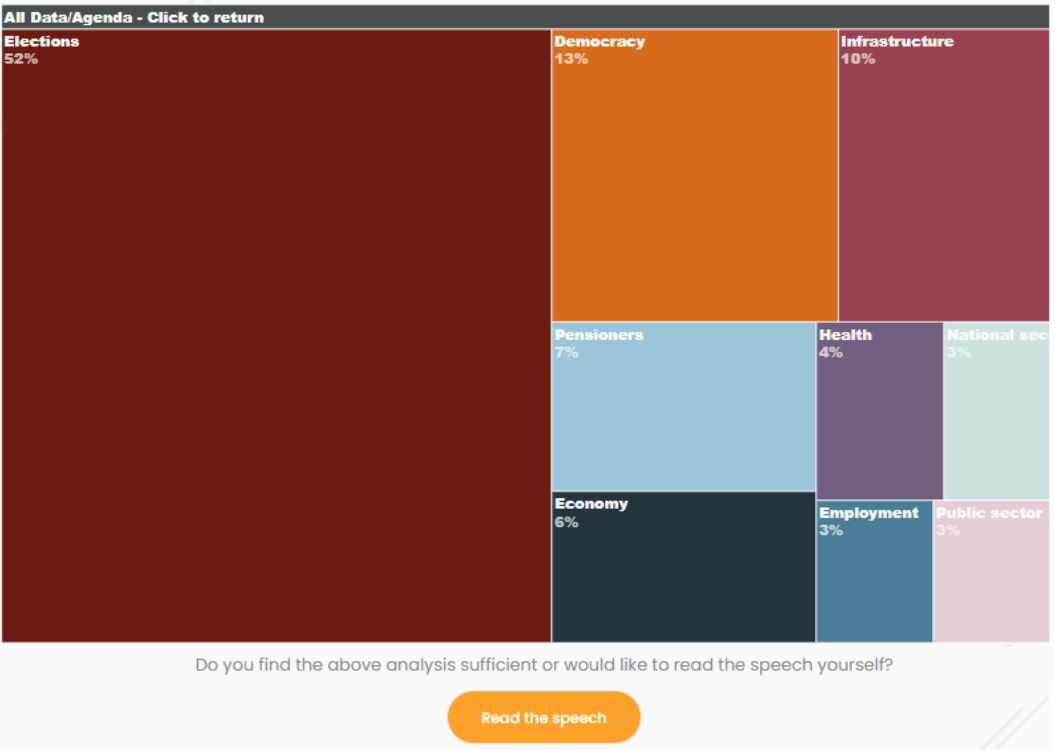}}\hfill
    \subfloat[\textbf{I2}-Comparative polarization accuracy assessment]{\label{sfig:task_2}\includegraphics[width=.49\textwidth, height=6cm]{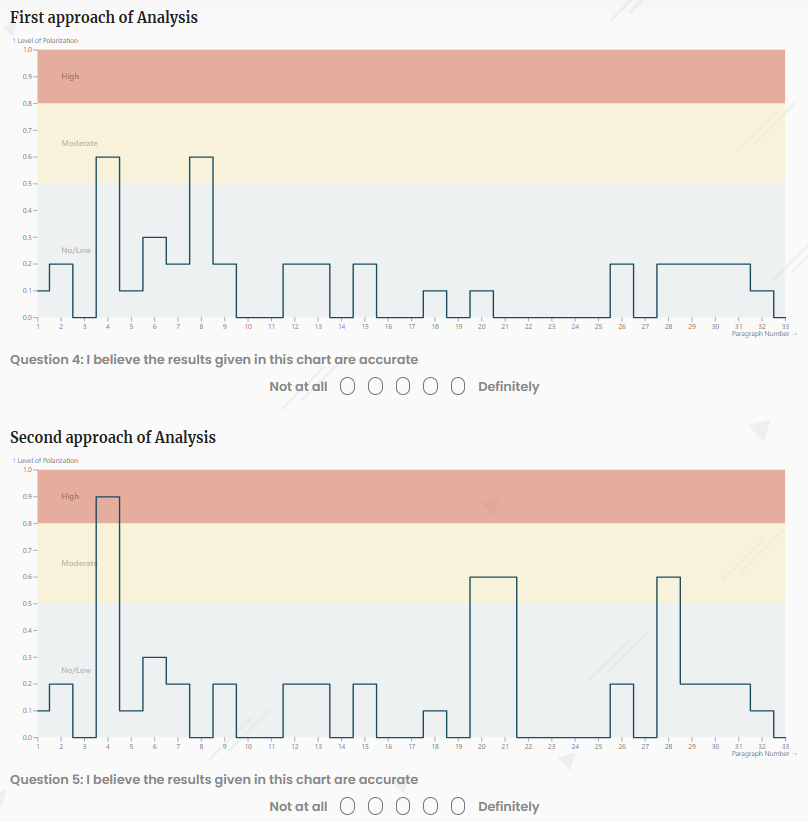}}\\
    \subfloat[\textbf{I3}-Fully-aware polarization accuracy assessment]{\label{sfig:task_3}\includegraphics[width=.49\textwidth, height=5cm]{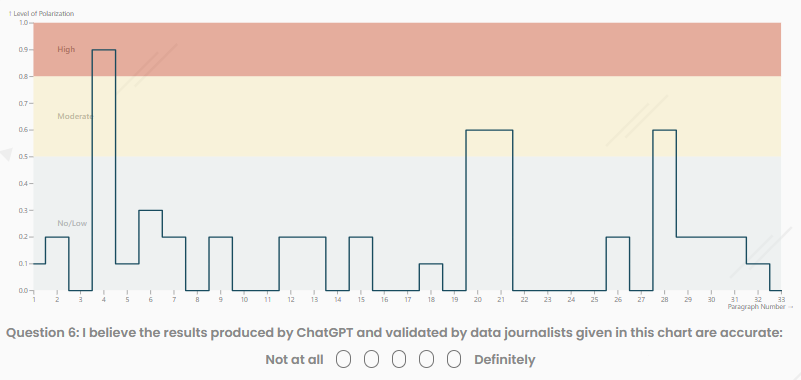}}\hfill
    \subfloat[\textbf{I4}-Sentiment opaqueness impact detection]{\label{sfig:task_4}\includegraphics[width=.49\textwidth, height=5cm]{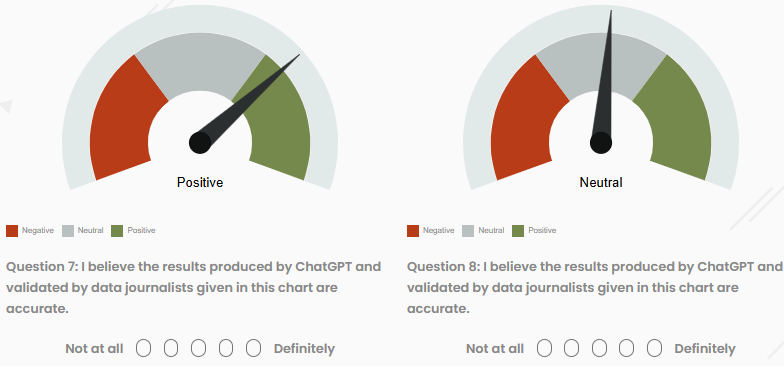}}\\
    \caption{The visualizations adopted in the demo version of the political discourse tool, along with the integrated question-answer fields for data collection.}
    \Description{The figure contains four sub-figures demonstrating the visualizations used in the demo tool. Sub-figure (a) shows a tree-map with the topics identified as used by political leaders in their speeches. Sub-figure (b) shows two step charts with the polarization levels identified in a speech using two different analysis approaches. Sub-figure (c) shows a step chart with the polarization levels identified in a speech. Sub-figure (d) shows two speedometers indicating the detected sentiment as positive, neutral, or negative.}
    \label{fig:demo_tool}
\end{figure*}

\end{document}